\documentstyle[prl,aps,twocolumn]{revtex}

\begin{document}

\preprint{To be published in Physical Review Letters}

\draft

\title{Conditional Quantum Dynamics and Logic Gates}

\author{Adriano Barenco, David Deutsch and Artur Ekert}
\address{Clarendon Laboratory, Physics Department, University of
  Oxford, Parks Road, Oxford OX1 3PU, United~Kingdom}
\author{Richard Jozsa}
\address{School of Mathematics \& Statistics, University of Plymouth,
Plymouth PL4 8AA, United~Kingdom}

\maketitle

\begin{abstract}
  Quantum logic gates provide fundamental examples of conditional
  quantum dynamics. They could form the building blocks of general
  quantum information processing systems which have recently been
  shown to have many interesting non--classical properties. We
  describe a simple quantum logic gate, the quantum controlled--NOT,
  and analyse some of its applications.  We discuss two possible
  physical realisations of the gate; one based on Ramsey atomic
  interferometry and the other on the selective driving of optical
  resonances of two subsystems undergoing a dipole--dipole
  interaction.
\end{abstract}

\begin{center} {PACS: 03.65.Bz, 03.75.Dg, 89.70.+c}
\end{center}

The fact that quantum--mechanical processes in principle allow new
types of information processing has been known for almost a
decade~\cite{Deutsch1,RF}. Bennett and Wiesner have shown that the
capacity of quantum channels can be doubled~\cite{BW} and recent
progress in quantum complexity theory~\cite{Complex} indicates that
the computational power of quantum computers exceeds that of Turing
machines. Hence the experimental realisation of such processes is a
most interesting issue. In this paper we concentrate on the basic
constituents of any quantum information processing device, namely
quantum logic gates. We wish to stress the apperance of a {\em
  conditional quantum dynamics}, in which one subsystem undergoes a
coherent evolution that depends on the quantum state of another
subsystem. The unitary evolution operator for the combined system has
the form

\begin{equation}
  U = |0\rangle \langle 0| \otimes U_0 + |0\rangle \langle 0|\otimes
  U_1 + \ldots + |k\rangle \langle k|\otimes U_k,
\end{equation}
where the projectors refer to quantum states of the {\em control}
subsystem and the unitary operations $U_i$ are performed on the {\em
  target} subsystem.  The simplest non--trivial operation of this sort
is the {\em quantum controlled--NOT}. We describe this gate, analyse
some of its applications and discuss physical realisations.

The {\em classical} controlled--NOT gate is a reversible logic gate
operating on two bits $\epsilon_1$ and $\epsilon_2$; $\epsilon_1$ is
called the control bit and $\epsilon_2$ the target bit. The value of
$\epsilon_2$ is negated if $\epsilon_1 =1$, otherwise $\epsilon_2$ is
left unchanged. In both cases the control bit $\epsilon_1$ remains
unchanged. We define the {\em quantum} controlled--NOT gate ${\cal
  C}_{12}$ as that which effects the unitary operation on two qubits
(two--state quantum systems), which in a chosen orthonormal basis
$\{|0\rangle,\, |1\rangle\}$ in ${\cal H}_2$ reproduces the
controlled--NOT operation:

\begin{equation} \label{c12} |\epsilon_1\rangle|\epsilon_2\rangle
  \stackrel{{\cal C}_{12} }{\longrightarrow}
  |\epsilon_1\rangle|\epsilon_1 \oplus\epsilon_2\rangle,
\end{equation}

where $\oplus$ denotes addition modulo 2. Here and in the following
the first subscript of ${\cal C}_{ij}$ always refers to the control
bit and the second to the target bit. Thus for example ${\cal C}_{21}$
performs the unitary operation defined by:
\begin{equation} |\epsilon_1\rangle|\epsilon_2\rangle
  \stackrel{{\cal C}_{21} }{\longrightarrow}
  |\epsilon_1\oplus\epsilon_2\rangle|\epsilon_2\rangle.
\end{equation}
The quantum controlled--NOT must be distinguished from the classical
controlled--NOT which is performable on existing computers. The
quantum controlled--NOT is a coherent operation on quantum states of
the two qubits.  The unitary operation defined by (\ref{c12}) is not
the only one which reproduces the classical controlled--NOT on the
computation basis states $|0\rangle$ and $|1\rangle$.  We may
introduce extra phases and the most general such quantum operation is:
\begin{equation}
  |\epsilon_1\rangle|\epsilon_2\rangle \longrightarrow \exp
  (i\theta_{\epsilon_1 \epsilon_2 })|\epsilon_1\rangle |\epsilon_1
  \oplus \epsilon_2\rangle.
\end{equation}
This phase would be irrelevant to classical operations but gives rise
to a family of inequivalent quantum gates.

Equations (2) or (3) define the gate ${\cal C}_{12}$ with respect to a
specific basis, the computation basis $\{|0\rangle,|1\rangle\}$. It is
also useful to consider generalisations of ${\cal C}_{12}$ which have
the analogous effect on the control and target bits in bases that are
different from the computation basis and possibly from each other. For
example ${\cal C}_{12}$ with respect to the basis
$\{\frac{1}{\sqrt{2}}(|0\rangle\pm|1\rangle)\}$ (for both qubits) is
easily shown to be identical to ${\cal C}_{21}$ with respect to the
basis $\{|0\rangle, |1\rangle\}$ {\em i.e.\/} the roles of the qubits
are reversed by this simple change of basis.  In the following we
always use the computation basis unless otherwise stated.

The quantum controlled--NOT gate has a variety of interesting
properties and applications:

(1) ${\cal C}_{12} $ transforms superpositions into entanglements
\begin{equation} C_{12}:\; (a|0\rangle+b|1\rangle)|0\rangle
  \longleftrightarrow a|0\rangle|0\rangle + b|1\rangle|1\rangle.
\end{equation} Thus it acts as a {\em measurement gate} because if the
target bit $\epsilon_2$ is initially in state $|0\rangle$ then this
bit together with the gate amount to an apparatus that performs a
perfectly accurate non--perturbing (quantum non--demolition ~\cite{qnd})
measurement of $\epsilon_1$.

(2) This transformation of superpositions into entanglements can be
reversed by applying the same controlled--NOT operation again. Hence
it can be used to implement the so--called {\em Bell
  measurement}~\cite{BMR92} on the two bits by disentangling the Bell
states. From the four Bell states we get four product states:
\begin{eqnarray} C_{12} \frac{1}{\sqrt
    2}(|0\rangle|0\rangle\pm|1\rangle|1\rangle) &=& \frac{1}{\sqrt
    2}(|0\rangle\pm|1\rangle) |0\rangle,\\ C_{12} \frac{1}{\sqrt
    2}(|0\rangle|1\rangle\pm|1\rangle|0\rangle) &=& \frac{1}{\sqrt
    2}(|0\rangle\pm|1\rangle) |1\rangle.
\end{eqnarray}
Thus the Bell measurement on the two qubits is effected by a sequence
of two independent two--dimensional measurements: in the computation
basis for the target qubit and in the basis $\{\frac{1}{\sqrt
  2}(|0\rangle\pm|1\rangle)\}$ for the control qubit. The realisation
of the Bell measurement is the main obstacle to the practical
implementation of quantum teleportation~\cite{teleport} and dense
quantum coding~\cite{BW}.

(3) Quantum state swapping can be achieved by cascading three quantum
controlled--NOT gates:
\begin{equation}
  C_{12}C_{21}C_{12} |\psi\rangle |\phi\rangle = |\phi\rangle
  |\psi\rangle,
\end{equation}
for arbitrary states $|\psi\rangle$ and $|\phi\rangle$ (see
also~\cite{RF1}).

(4) The quantum controlled--NOT gate may also be used to swap
distantly separated states in the presence of a channel carrying only
classical information. This is in contrast to the state swapping
described above which requires the gate to be applied to the two
states as inputs, so that they cannot be distantly separated at the
time. Suppose that Alice and Bob, distantly separated, have states
$|\alpha\rangle$ in ${\cal H}_0$ and $|\beta\rangle$ in ${\cal H}_5$
respectively which they wish to swap (the identities of the states are
assumed to be unknown to Alice and Bob). To achieve this they will
need, on a previous occasion when they were close together or had
access to a quantum communication channel, to have shared two pairs of
qubits, one in the state $\frac{1}{\sqrt{2}}(|0\rangle|0\rangle +
|1\rangle|1\rangle)$ in ${\cal H}_1 \otimes {\cal H}_3 $ and the other
in the same maximally entangled state in ${\cal H}_2 \otimes {\cal
  H}_4 $. States in ${\cal H}_0$,${\cal H}_1$, ${\cal H}_2$ are
localised near Alice and states in ${\cal H}_3$, ${\cal H}_4$, ${\cal
  H}_5$ are localised near Bob. Let ${\cal M}$ denote a complete
measurement in the computational basis $\{|0\rangle,|1\rangle\}$.

To swap $|\alpha\rangle$ and $|\beta\rangle$ Alice and Bob carry out
the following protocol. {\em Step 1}: Alice performs ${\cal C}_{10}$
and then ${\cal C}_{02}$ while Bob performs ${\cal C}_{54}$ and then
${\cal C}_{35}$.  {\em Step 2}: Alice measures ${\cal M}$ in ${\cal
  H}_2$ and Bob measures ${\cal M}$ in ${\cal H}_4$. Each communicates
the result (one bit of information) to the other participant. If the
results are the same, go to step 3. If the results are different,
Alice and Bob negate all bits in their possession {\em i.e.\/} apply
the unitary operation \[\left(
\begin{array}{cc}
 0 & 1 \\
 1 & 0
 \end{array} \right) \] to each particle. {\em Step 3}: Alice applies the
rotation
 \[ \frac{1}{\sqrt{2}} \left( \begin{array}{lr}
 1 & 1 \\
 1 & -1
 \end{array} \right) \]
 in ${\cal H}_1$ and Bob does the same in ${\cal H}_3$.  {\em Step 4}:
 Alice performs the measurement ${\cal M}$ in ${\cal H}_1$ and Bob
 performs it in ${\cal H}_3$. Each communicates the result to the
 other. If the results agree then the states will have been swapped.
 If the results differ then Alice applies the unitary transformation
\[ \left( \begin{array}{lr}
 1 & 0 \\
 0 & -1 \end{array} \right) \] to ${\cal H}_0$ and Bob does the same to
${\cal H}_5$ after which the states will have been swapped. A related
process has been described by Vaidman\cite{vaidman}.

It is interesting to compare the protocol to {\em quantum
  teleportation} \cite{teleport}, in which Alice and Bob initially
share one maximally entangled pair and Alice is able to transfer an
arbitrary state $|\xi\rangle$ to Bob by sending him only two classical
bits of information.  Thus using the same resources as in our protocol
{\em viz.} sharing two entangled pairs and each participant sending
two bits to the other, we may alternatively swap the states
$|\alpha\rangle$ and $|\beta\rangle$ by performing two teleportations
(for the two directions of transfer.) However, the process above
cannot be separated into two successive transfers. The remarkable
feature of all these processes is that in the presence of shared
entanglement an arbitrary state $|\xi\rangle$ may be transferred as a
result of sending only a few bits of classical information despite the
fact that $|\xi\rangle$ depends on two continuous parameters
corresponding to an {\em infinite} amount of classical information.

The quantum controlled--NOT gate is not a universal gate. However,
together with relatively trivial single--qubit operations it forms an
adequate set of quantum gates {\em i.e.\/} the set from which any
quantum gate may be built~\cite{torino94}. Thus the conditional
dynamics of the quantum controlled--NOT type would in realistic
technologies be sufficient to construct any quantum information
processing device.  Universal two--qubit quantum gates based on
similarly controlled dynamics are described in~\cite{Ugates}.

In the following we outline two possible experimental realisations of
the quantum controlled--NOT gate. We do not wish to suggest that these
particular technologies are destined to yield practicable devices.
They do, however, serve to illustrate physical considerations that
would bear on the building of such devices in any technology.

The first technology is that of Ramsey atomic
interferometry~\cite{Ramsey,DZBRH94,Haroche1,qedrev}, the second is
based on the selective driving of optical resonances of two qubits
undergoing a dipole--dipole interaction~\cite{OTM}.

In the Ramsey atomic interferometry method the target qubit is an atom
with two circular Rydberg states $|\epsilon_2\rangle$, where
$\epsilon_2 =0,1$; the control qubit is the quantized electromagnetic
field in a high--$Q$ cavity $C$. The field in the cavity contains at
most one photon of a particular mode so it can be viewed as a two
state system with the vacuum state $|0\rangle$, and the one--photon
state $|1\rangle$ as the basis. The cavity $C$ is sandwiched between
two auxiliary microwave cavities $R_1$ and $R_2$ in which classical
microwave fields produce $\pi/2$ rotations of the Bloch vector of an
atom passing through at a given speed,

\begin{equation}
  |\epsilon_1\rangle_{\mbox {\scriptsize field}}
  |\epsilon_2\rangle_{\mbox{\scriptsize atom}} \longrightarrow
  |\epsilon_1\rangle_{\mbox {\scriptsize field}}\;
\frac{1}{\sqrt 2}
 ( |\epsilon_2\rangle +
 (-1)^{\epsilon_2}e^{i\alpha} |1-\epsilon_2\rangle)_{\mbox{\scriptsize
 atom}}.
\end{equation}

where the phase factor $\alpha$ is different for the two cavities
$R_1$ and $R_2$. In the central cavity $C$, a dispersive interaction
with the quantized field introduces phase shifts which depend on the
state of the atom $|\epsilon_2\rangle$ and on the number of photons in
the cavity $|\epsilon_1\rangle$. The interaction conserves the number
of photons in the cavity.

\begin{equation}
  |\epsilon_1\rangle_{\mbox {\scriptsize field}}
  |\epsilon_2\rangle_{\mbox{\scriptsize atom}} \longrightarrow
  \exp (i
 (-1)^{1-\epsilon_2} (\epsilon_1+\epsilon_2) \/ \theta\/)
  |\epsilon_1\rangle_{\mbox {\scriptsize field}}
  |\epsilon_2\rangle_{\mbox{\scriptsize atom}},
\end{equation}

where $\theta$, the phase shift per photon, can be tuned to be $\pi$
(\/ $\theta$ depends on the time taken for the atom to cross $C$ and
the atom--field detuning).

The overall process can be viewed as a sequence: half--flipping in
$R_1$, phase shifts in $C$, and half--flipping in $R_2$. Depending on
the phase shifts the second half--flipping can either put the atom
back into its initial state or flip it completely into the orthogonal
state. The whole interferometer can be adjusted so that when the atom
passes successively through the cavities $R_1$, $C$, and $R_2$ the two
qubits, {\em i.e.\/} the field and the atom, undergo the
transformation

\begin{equation}
  |\epsilon_1\rangle_{\mbox {\scriptsize field}}
  |\epsilon_2\rangle_{\mbox{\scriptsize atom}} \longrightarrow
  |\epsilon_1\rangle_{\mbox {\scriptsize field}}  |\epsilon_1
\oplus
 \epsilon_2\rangle_{\mbox{\scriptsize atom}}.
\end{equation}

The state of the field in $C$ can also be transfered from (and to) an
auxiliary Rydberg atom tuned to the resonant frequency of the cavity
so that it undergoes a resonant rather than a dispersive interaction
in $C$.  This process allows the creation of a gate acting on two
qubits of the same type, {\em i.e.\/} two Rydberg atoms rather than a field
and an atom.  Davidowich {\em et al.}~\cite{DZBRH94} have shown how to
use the Ramsey interferometry for quantum teleportation. Their
experimental setup effectively contains conditional dynamics of the
type we have been discussing, which has much wider application in
quantum information processing than merely quantum teleportation.  The
practical realisation of the quantum controlled--NOT gate can be
achieved with a modification of the experiments as described
in~\cite{DZBRH94,Haroche1,qedrev}. The typical resonant frequency
would be of the order $\sim 2\times 10^{10}$~Hz, the atom--field
interaction time in the cavity $\sim 3\times 10^{-5}$~s, and the
cavity field lifetime can be made as long as $\sim 0.5$~s.

The most difficult part of the experimental realisation is likely to
be the preparation of a single atom. This is usually done by preparing
an atomic beam with a very low probability of finding a single atom in
the beam; consequently finding two atoms in the beam is even less
probable.  From our point of view the drawback of this method is that
it forces a trade--off between the probability that precisely one atom
(as required) interacted with the field on a given run, and the
reliability of the gate.  Although our example has focused on
microwave cavities, experimental realisations in the optical regime
could also be considered~\cite{qedrev}.

Our second proposal for the implementation of the quantum
controlled--NOT gate relies on the dipole--dipole interaction between
two qubits. For the purpose of this model the qubits could be either
magnetic dipoles, e.g. nuclear spins in external magnetic fields, or
electric dipoles, e.g. single--electron quantum dots in static electric
fields. Here we describe the model based on interacting quantum dots,
but mathematically the two cases are isomorphic.

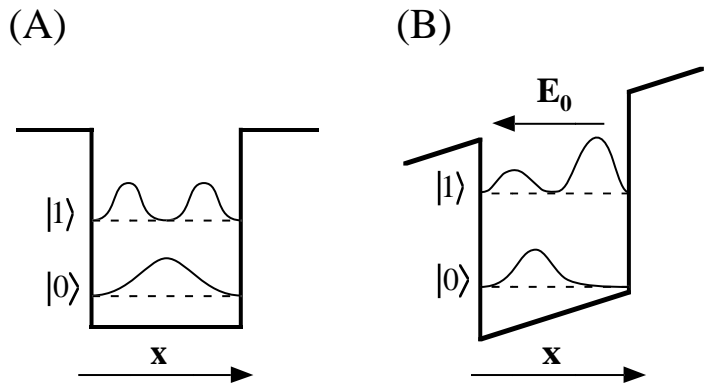
\begin{figure}[h]
\caption[foo2]{\small Charge density in the quantum well in the direction
  ${\bf x}$ of the applied field. A dipole moment is induced when the
  electric field is turned on (B), but is zero without the electric
  field (A). }
\label{wells}
\end{figure}

Two single--electron quantum dots separated by a distance $R$ are
embedded in a semiconductor. Let us consider the ground state and the
first excited state of each dot as computation basis states
$|0\rangle$ and $|1\rangle$.  The first quantum dot, with resonant
frequency $\omega_1$, will act as the control qubit and the second
one, with resonant frequency $\omega_2$, as the target qubit. In the
presence of an external static electric field, which can be turned on
and off adiabatically in order to avoid transitions between the
levels, the charge distribution in the ground state of each dot is
shifted in the direction of the field whilst in the first excited
state the charge distribution is shifted in the opposite direction
(the {\em quantum--confined Stark effect})\cite{MCS86}, see
Fig.~\ref{wells}. In the simple model in which the state of the qubit
is encoded by a single electron per quantum dot, we can choose
coordinates in which the dipole moments in states $|0\rangle$ and
$|1\rangle$ are $\pm d_i$, where $i=1,2$ refers to the control and
target dot respectively. For the sake of clarity, we are presenting
the idea using a slightly simplified model. A more elaborate model
would take into account holes in the valence band of the
semiconductors. The state of a qubit would be determined by excitons
of different energies.

The electric field from the electron in the first quantum dot may
shift the energy levels in the second one (and vice versa), but to a
good approximation it does not cause transitions. That is because the
total Hamiltonian
\begin{equation}
  \hat H = \hat H_1 + \hat H_2 + \hat V_{12}
\end{equation}
is dominated by a dipole--dipole interaction term $\hat V_{12}$ that is
diagonal in the four--dimensional state space spanned by eigenstates
$\{ |\epsilon_1\rangle, |\epsilon_2\rangle\}$ of the free Hamiltonian
$\hat H_1+\hat H_2$, where $\epsilon_1$ and $\epsilon_2$ range over 0
and 1. Specifically,
\begin{equation}  (\hat H_1 + \hat H_2)  |\epsilon_1\rangle
  |\epsilon_2\rangle = \hbar (\epsilon_1\omega_1 + \epsilon_2\omega_2)
  |\epsilon_1\rangle |\epsilon_2\rangle,
\end{equation}
and
\begin{equation}
  \hat V_{12} |\epsilon_1\rangle |\epsilon_2\rangle = (-1)^{\epsilon_1
    +\epsilon_2}\hbar\bar \omega |\epsilon_1\rangle
  |\epsilon_2\rangle,
\end{equation}
where
\begin{equation}
  \bar\omega = -\frac{d_1d_2}{4\pi\epsilon_0 R^3}.
\end{equation}

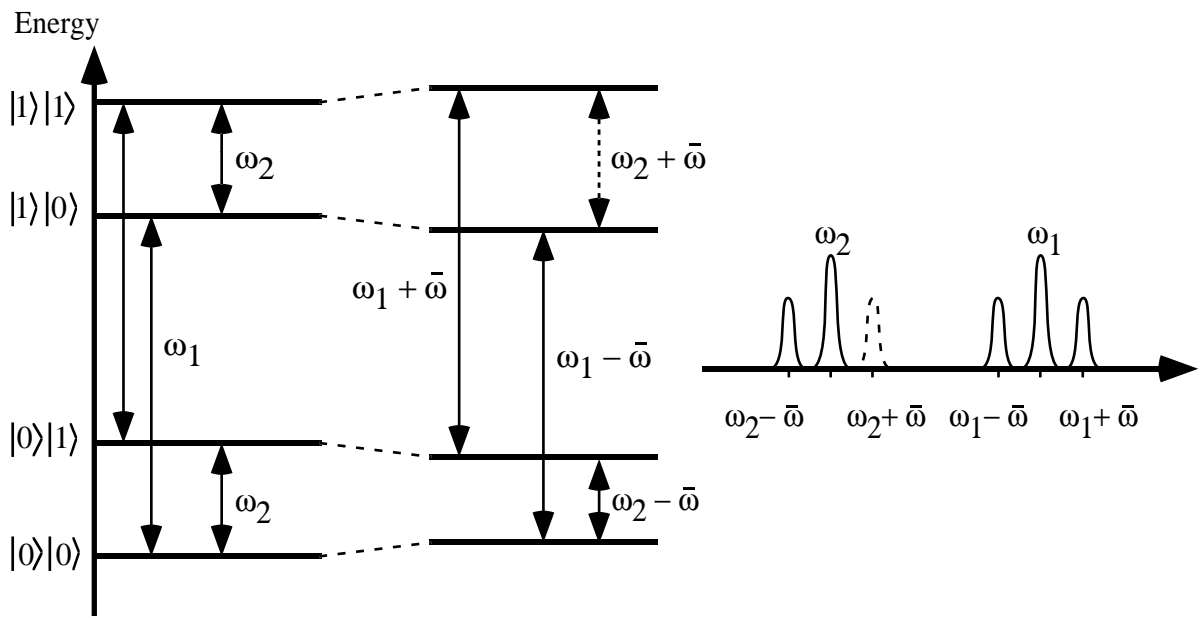
\begin{figure}[h]
\caption[foo1]{\small (a) Energy levels of two quantum dots
  without and with the coupling induced by the presence of a static
  electric field ${\bf E}_0$. (b) Resonance spectrum of the two
  quantum dots.  The dotted line shows the wavelength for which the
  two dots act as a controlled--NOT gate, with the first dot being the
  control qubit and the second the target qubit. }
\label{shifts}
\end{figure}

As shown in Fig.~\ref{shifts}, it follows that due to the
dipole--dipole interaction the resonant frequency for transitions
between the states $|0\rangle$ and $|1\rangle$ of one dot {\em depends
  on the neighbouring dot's state}. This is the conditional quantum
dynamics we are seeking. The resonant frequency for the first dot
becomes $\omega_1\pm\bar\omega$ according as the second dot is in
state $|0\rangle$ or $|1\rangle$ respectively. Similarly the second
dot's resonant frequency becomes $\omega_2\pm\bar\omega$, depending on
the state of the first dot. Thus a $\pi$--pulse at frequency
$\omega_2+\bar\omega$ causes the transition $|0\rangle \leftrightarrow
|1\rangle$ in the second dot if and only if the first dot is in state
$|1\rangle$.

For such processes to be useful for quantum information processing,
the decoherence time must be greater than the time scale of the
optical interaction (see for example~\cite{Unruh}). The decoherence
time depends partly on the modification of the confining potential due
to phononic excitations. There is also a quantum electrodynamic
contribution due to coupling to the vacuum modes. For resonant
frequencies in the infrared regime, this can be estimated at $\sim
10^{-6}$~s. Impurities and thermal vibration (phonons) can reduce the
lifetime further to $\sim 10^{-9}$~s or even worse, but in principle
their effects can be minimized by a more precise fabrication
technology and by cooling the crystal.  The optical interaction
time--scale can be approximated by the length of the $\pi$--pulse
($\sim 10^{-9}$~s).  The length of the pulse is not so much restricted
by the current technology as by the requirement for the $\pi$--pulse
to be monochromatic and selective enough; this restricts the length of
the pulse to being greater than the inverse of the pulse carrier
frequency and the inverse of the dipole--dipole interaction coupling
constant ($1/\bar\omega \sim 10^{-12}$~s in our model). This model may
be more difficult to implement than the one based on the Ramsey
atomic interferometry, but once it is implemented it is likely allow
for quantum gates to be integrated more easily into complex quantum
circuits, as required for more general quantum information processing.

\section*{Acknowledgements} The authors thank B.G.~Englert, S.~Haroche,
H.J.~Kimble, H.~Mabuchi, G.~Mahler, J--M.~Raimond, H.~Walther for
discussions and comments. This work has been partially supported by
the NIST Advanced Technology Program. A.~B. gratefully acknowldeges
the financial support of the Berrow's Fund at Lincoln College
(Oxford). Work by A.~E. is supported by the Royal Society, London.

\section*{Captions}

\begin{thebibliography}{99}

\bibitem{Deutsch1} D.~Deutsch, Proc.~R.~Soc.~London~A {\bf 400}, 97
  (1985).

\bibitem{RF} R.~Feynman, Int.~J.~Theor.~Phys. {\bf 21}, 467 (1982).

\bibitem{BW} C.H.~Bennett and S.J.~Wiesner, Phys.~Rev.~Lett. {\bf 69},
  2881 (1992).

\bibitem{Complex} D.~Deutsch and R.~Jozsa, Proc.~R.~Soc.~Lond.~A {\bf
    439}, 553 (1992); A.~Berthiaume and G.~Brassard, J.~Mod.~Opt.
  {\bf 41}, 2521 (1994); R.~Josza, Proc.~R.~Soc.~Lond.~A {\bf 435} 563
  (1991); D.~Simon {\em Proc. 35th Ann. Symp. Foundations of Computer
    Science\/}, IEEE Press, 116 (1994); E.~Bernstein and U.~Vazirani,
  {\em Proc.~25th~ACM Symp. on Theory of Computation\/},11 (1993);
  P.W.~Shor, {\em Proc. 35th Ann. Symp. Foundations of Computer
    Science\/}, IEEE Press (1994).

\bibitem{qnd} V.B.~Braginsky, Yu.~I.~Vorontsov, and F.~Ya.~Khalili,
  Zh.~Eksp.~Theo.~Fiz. {\bf 73}, 1340 [Sov. Phys. JETP {\bf 46}, 705
  (1977)].

\bibitem{BMR92} S.L. Braunstein, A. Mann, and M. Revzen, Phys. Rev.
  Lett. {\bf 68}, 3259 (1992).

\bibitem{teleport} C.H.~Bennett, G.~Brassard, C.~Cr\'epeau, R.~Jozsa,
  A.~Peres, and W.K.~Wootters, Phys.~Rev.~Lett. {\bf 70}, 1895 (1993).

\bibitem{RF1} R.~Feynman, Opt. News {\bf 11}, 11 (1985).

\bibitem{vaidman} L.~Vaidman, Phys.~Rev.~A {\bf 49}, 1473 (1994).

\bibitem{torino94} A.~Barenco, C.H~Bennett, R~.Cleve, D.~DiVincenzo,
  N.~Margolus, P.~Shor, T.~Sleator, J.~Smolin and H.~Weinfurter ({\em
    preprint\/}).

\bibitem{Ugates} D.~Deutsch, Proc.~R.~Soc.~Lond.~A {\bf 425}, 73
  (1989); A.~Barenco, Phys.~Rev.~Lett (submitted); T~Sleator and
  H.~Weinfurter Phys.~Rev.~Lett. (submitted); D.~Deutsch, A.~Barenco,
  and A.~Ekert, Proc.~R.~Soc.~Lond.~A (submitted).

\bibitem{Ramsey} N.F.~Ramsey {\em Molecular Beams.} Oxford University
  Press (1985).

\bibitem{DZBRH94} L.~Davidovich, N.~Zagury, M.~Brune, J--M.~Raimond,
  and S.~Haroche, Phys.~Rev.~A, {\bf 50}, R895 (1994); T.~Sleator and
  H.~Weinfurter (private communication).

\bibitem{Haroche1} M.~Brune, P.~Nussenzveig, F.~Schmidt--Kaler,
  F.Bernardot, A.~Maali, J--M.~Raimond, and S.~Haroche,
  Phys.~Rev.~Lett. (1994).

\bibitem{qedrev} Advances in Atomic, Molecular and Optical Physics,
  Supplement 2 (vol. entitled {\em Cavity QED\/}), edited by P. Berman
  (Academic Press, 1994).

\bibitem{OTM} K.~Obermayer, W.G.~Teich, and G.~Mahler, Phys.~Rev.~B
  {\bf 37}, 8096 (1988); W.G.~Teich, K.~Obermayer, and G.~Mahler, {\em
    ibid.} p.8111; W.G.~Teich and G.~Mahler, Phys.~Rev.~A {\bf 45},
  3300 (1992); see also S.~Lloyd, Science {\bf 261}, 1569 (1993).

\bibitem{MCS86} D.A.B. Miller, D.S. Chemla, and S. Schmitt--Rink,
  Phys.~Rev.~B {\bf 33}, 6976 (1986).

\bibitem{Unruh} W.~Unruh, Phys.~Rev.~A (submitted).

\end{thebibliography}
\end{document}